\documentclass[10pt,letterpaper]{article}

\usepackage{cogsci}
\usepackage{pslatex}
\usepackage{apacite}
\usepackage{amsmath}
\usepackage{amsfonts}
\usepackage{amssymb}
\usepackage{graphicx}
\usepackage[round]{natbib}
\usepackage{color}
\usepackage{subfig}
\usepackage{dblfloatfix}
\title{Over-representation of Extreme Events in Decision-Making:\\ A Rational Metacognitive Account}

\author{{\large \bf Ardavan S. Nobandegani$^{1,3}$, Kevin~da~Silva~Castanheira$^{3}$, A.~Ross~Otto$^{3}$, \& Thomas~R.~Shultz$^{2,3}$}\\
\{ardavan.salehinobandegani, kevin.dasilvacastanheira\}@mail.mcgill.ca\\
\vspace*{-2pt}\{ross.otto, thomas.shultz\}@mcgill.ca\vspace*{3pt}\\
\vspace*{-1pt}\small{$^{1}$Department of Electrical \& Computer Engineering, McGill University}\\
\vspace*{-1pt}\small{$^{2}$School of Computer Science, McGill University}\\
\vspace*{-1pt}\small{$^{3}$Department of Psychology, McGill University}}

\begin{document}
\maketitle

\begin{abstract}
The Availability bias, manifested in the over-representation of extreme eventualities in decision-making, is a well-known cognitive bias, and is generally taken as evidence of human irrationality. In this work, we present the first rational, metacognitive account of the Availability bias, formally articulated at Marr's algorithmic level of analysis. Concretely, we present a normative, metacognitive model of how a cognitive system should over-represent extreme eventualities, depending on the amount of time available at its disposal for decision-making. Our model also accounts for two well-known framing effects in human decision-making under risk---the fourfold pattern of risk preferences in outcome probability (Tversky \& Kahneman, 1992) and in outcome magnitude (Markovitz, 1952)---thereby providing the first metacognitively-rational basis for those effects. Empirical evidence, furthermore, confirms an important prediction of our model. Surprisingly, our model is unimaginably robust with respect to its focal parameter. We discuss the implications of our work for studies on human decision-making, and conclude by presenting a counterintuitive prediction of our model, which, if confirmed, would have intriguing implications for human decision-making under risk. To our knowledge, our model is the first {metacognitive}, resource-rational process model of cognitive biases in decision-making.

\end{abstract}

\section{Introduction} 
Which one comes to your mind easier? {The most horrible car crash of your life, or the event of driving home safely on an ordinary day?} Among the great many cognitive biases documented in the literature, the Availability bias (Tversky \& Kahneman, 1972) is a notable one: people overestimate the probability of events that easily come to mind. A number of notable effects can be explained by this cognitive bias: people's overestimation of the frequency of extreme events like an earthquake (Lichtenstein, Slovic, Fischhoff, Layman, \& Combs, 1978) and people's overreaction to threats like terrorism (Lichtenstein et al., 1978; Rothman, Klein, \& Weinstein,
1996, Sunstein \& Zeckhauser, 2011). Neurobiological work shows that the strength of a memory is modulated by the salience of its positive or negative valance (Cruciani et al., 2011), thereby providing a possible explanation of the Availability bias.

Recently, Lieder, Griffiths, and Hsu (2014, 2017) proposed a boundedly-optimal, rational process model of this bias which can explain a wide range of findings in the human-decision making literature. Drawing on the \emph{importance sampling paradigm}, their account aimed to minimize the mean squared error (MSE) of an expected utility estimator, as a well-established and normatively-justified measure of quality of an estimator (Poor, 2013). Since the variance of the estimator is the asymptotically-dominant  term in the MSE (i.e., for large sample size, variance becomes an accurate proxy for MSE), Lieder et al.~(2014, 2017) suggested that people adopt the following importance distribution (as the importance distribution minimizing the variance): 
\begin{eqnarray}
\label{eq_lieder_q}
q(o)\propto p(o)|u(o)-\mathbb{E}_p[u(o)]|,
\end{eqnarray}
for mental simulations of events. In~(\ref{eq_lieder_q}), $o$ denotes an arbitrary event, $p$ the objective probability of event $o$, $u(o)$ the utility of event $o$, $q$ the probability distribution one adopts for their mental simulations (i.e., the subjective probability of event $o$), and, finally, $\mathbb{E}_p[\cdot]$ the expectation with respect to $p$.

Note that the expression in (\ref{eq_lieder_q}) does not depend on the number of samples one gets to draw before making their decision (denoted by $s$). In that light, Lieder et al.'s (2014, 2017) account implies that \emph{time availability}, i.e., the amount of time a decision-maker has for making a decision, should have no implications on what importance distribution $q$ one adopt. While a cognitively-rational agent is ignorant about adapting their importance distribution $q$ based on time availability, a \emph{meta}cognitively-rational agent plausibly considers that in their choice of $q$. That is, the metacognitively-rational agent chooses, among all $q$'s, the one which is normatively-justified based on time availability considerations---this essentially makes it a strategy selection task guided by time availability. In agreement with this view, a large body of psychological work on decision-making suggests that (1) people evoke different strategies for decision making under time pressure vs.~no time pressure condition, and (2) people adapt their strategies in accord with time availability (see e.g.,~Svenson \& Maule, 1993; Svenson, 1993). 

In this work, we present the first normative, metacognitive model of how an agent should over-represent extreme eventualities, depending on the amount of time available at their disposal for
decision making. Concretely, our work serves as a rational, meta-level model for the work by Lieder et al.~(2017, 2014). More specifically, the importance distribution suggested by Lieder et al.~(2017, 2014) naturally follows from our metacognitive account, when s is large (i.e. for large sample size regime). In contrast to Lieder et al.~(2017, 2014), our meta-level account also specifies how a decision-maker should rationally choose their importance distribution when they can only afford to collect merely very few samples (i.e.~when making decision under extremely high time pressure).\footnote{The optimality of Lieder et al.'s (2017, 2014) model hinges on the number of samples $s$ being large. When $s$ is small (i.e.~small sample size regime) Lieder et al.'s (2017, 2014) model is no longer optimal. Our model, however, remains rational for both small and large $s$'s.} Importantly, recent work has provided mounting evidence suggesting that people often use very few samples in probabilistic judgments and reasoning under uncertainty (e.g., Vul et al., 2014; Battaglia et al. 2013; Lake et al., 2017; Gershman, Horvitz, \& Tenenbaum, 2015; Hertwig \& Pleskac, 2010; Griffiths et al., 2012; Gershman, Vul, \& Tenenbaum, 2012; Bonawitz et al., 2014), elevating the importance of developing process models specifically directed at the small sample size regime. 

We show that our model can account for two well-known framing effects in human decision-making under risk: the fourfold pattern of risk preferences in outcome probability (Tversky \& Kahneman, 1992) and in outcome magnitude (Markovitz, 1952). Despite being often taken as strong evidence for human irrationality, we provide the first metacognitively-rational basis for these effects. Empirical evidence, furthermore, confirms an important prediction of our model:~over-representation of extreme events regardless of their frequencies. Our model also makes a counterintuitive (normative) prediction, which, if confirmed, would have surprising implications for human decision-making under risk.

\section{Proposed Model}
In this section, we formally present our metacognitively-rational model for the Availability bias (Tversky \& Kahneman, 1973). According to the expected utility theory (Von Neumann \& Morgenstern, 1944), an agent chooses an action $a$, with the highest expected utility
\begin{eqnarray}
\mathbb{E}[u(o)]=\int p(o|a)u(o)do,
\label{eq_exp_utility}
\end{eqnarray} 
where $p(o|a)$ denotes the distribution over outcomes $o$ resulting from taking action $a$, $u(o)$ the subjective utility associated to outcome $o$, and $\mathbb{E}[\cdot]$ the expectation operation. Since the computation of (\ref{eq_exp_utility}) is intractable in general, we assume that the agent estimates (\ref{eq_exp_utility}) using \emph{sampling methods} (Hammersley \&
Handscomb, 1964). Substantial neural and behavioral evidence supports this hypothesis (see e.g.~Fiser, Berkes, Orb´an, \& Lengyel, 2010; Vul, Goodman, Griffiths, \& Tenenbaum, 2014; Denison, Bonawitz, Gopnik, \& Griffiths, 2013; Griffiths \& Tenenbaum, 2006). Concretely, following Lieder et al.~(2014, 2017), we assume that the agent estimates~(\ref{eq_exp_utility}) using (self-normalized) importance sampling (Hammersley \&
Handscomb, 1964; Geweke, 1989), which is shown to have connections to both neural networks (Shi \& Griffiths, 2009) and cognitive process models (Shi, Griffiths, Feldman, \& Sanborn, 2010):
\begin{eqnarray}
\hat{E}=\dfrac{1}{\sum_{j=1}^s w_j} \sum_{i=1}^s w_i u(o_i), \quad \forall i:\ o_i\sim q,\ w_i=\dfrac{p(o_i)}{q(o_i)}.
\label{eq_norm_is_estimator}
\end{eqnarray}
In Eq.~(\ref{eq_norm_is_estimator}), $s$ denotes the total number of mental simulations performed by the agent, $o_i$ the $i^{th}$ mentally simulated outcome, $u(o_i)$ the utility of $o_i$, $p$ the objective probability of event $o_i$, $q$ the probability distribution the agent adopts for their mental simulations (i.e., the subjective probability of event $o_i$), and, $\hat{E}$ the (normalized) importance sampling estimator of $\mathbb{E}[u(o)]$ given in (\ref{eq_exp_utility}). 

The mean-squared error (MSE) of the estimator in (\ref{eq_norm_is_estimator}), as a standard normative measure of the quality of an estimator (Poor, 2013), can be decomposed as follows: $\mathbb{E}[(\hat{E}-\mathbb{E}[u(o)])^2]=(\text{Bias}[\hat{E}])^2+\text{Var}[\hat{E}]$, where
The bias $\text{Bias}[\hat{E}]$ and variance $\text{Var}[\hat{E}]$ of the estimator $\hat{E}$ can be approximated by (Zabaras, 2010):
\begin{eqnarray}
\text{Bias}[\hat{E}]\approx\dfrac{1}{s} \int \dfrac{p(o)^2}{q(o)}(\mathbb{E}_p[u(o)]-u(o))do,\\
\text{Var}[\hat{E}]\approx\dfrac{1}{s} \int \dfrac{p(o)^2}{q(o)}(\mathbb{E}_p[u(o)]-u(o))^2 do.
\end{eqnarray}

Under mild technical conditions, it can be shown that the rational importance distribution for minimizing the MSE of the estimator $\hat{E}$ is given by:
\begin{eqnarray}
q^\ast_{meta}\propto p(o)|u(o)-\mathbb{E}_p[u(o)]| \sqrt{\dfrac{1+|u(o)-\mathbb{E}_p[u(o)]|\sqrt{s}}{|u(o)-\mathbb{E}_p[u(o)]|\sqrt{s}}},
\label{eq_meta_rational_q}
\end{eqnarray}
where $p$ denotes the objective probability of event $o$, and $\mathbb{E}_p[\cdot]$ the expectation with respect to distribution $p$. We refer to $q^\ast_{meta}$ given in (\ref{eq_meta_rational_q}) as the metacognitively-rational importance distribution the agent should adopt for mental simulation of events for decision-making under uncertainty. For the derivation of the expression given in  (\ref{eq_meta_rational_q}), the reader is referred to Sec.~A-I of the Appendix.

Comparing expressions (\ref{eq_lieder_q}) and (\ref{eq_meta_rational_q}) reveals that the multiplicative factor $\textstyle\sqrt{\dfrac{1+|u(o)-\mathbb{E}_p[u(o)]|\sqrt{s}}{|u(o)-\mathbb{E}_p[u(o)]|\sqrt{s}}}$, which we term  \emph{metacognitive rationality factor} (MCRF), is what sets apart Lieder et al.'s (2014, 2017) cognitively-rational model (see Eq.~(\ref{eq_lieder_q})) from our metacognitively-rational model. In the remainder of this work, we show that MCRF plays a crucial role in accounting for two important framing effects in decision-making under risk. It is crucial to note that $q^\ast_{meta}$ takes into account the amount of time available for making a decision (i.e.,~time availability), as evidenced by expression (\ref{eq_meta_rational_q}) explicitly depending on the number of mental simulations $s$ performed by the agent.

Following Lieder et al.~(2017, 2014), and for ease of exposition, we assume $\mathbb{E}_p[u(o)]=0$ hereinafter.

\setcounter{figure}{1}
\begin{figure*}[bp]
    \centering
    \subfloat[]{{\includegraphics[trim = 5mm 90mm 10mm 90mm,clip,width=0.5\textwidth]{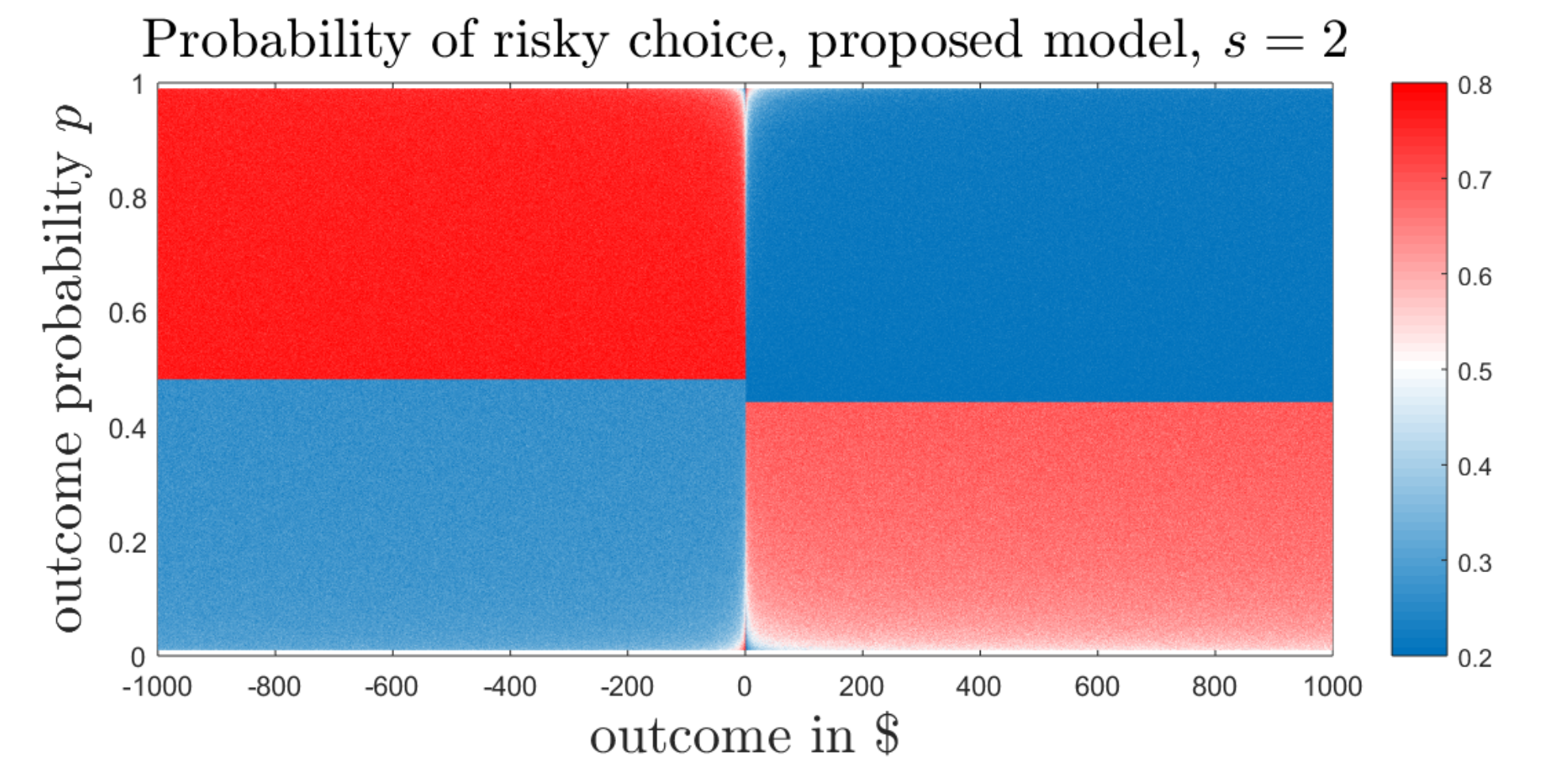} }}
    \hspace*{1pt}
    \subfloat[]{{\includegraphics[trim = 5mm 90mm 10mm 90mm,clip,width=0.5\textwidth]{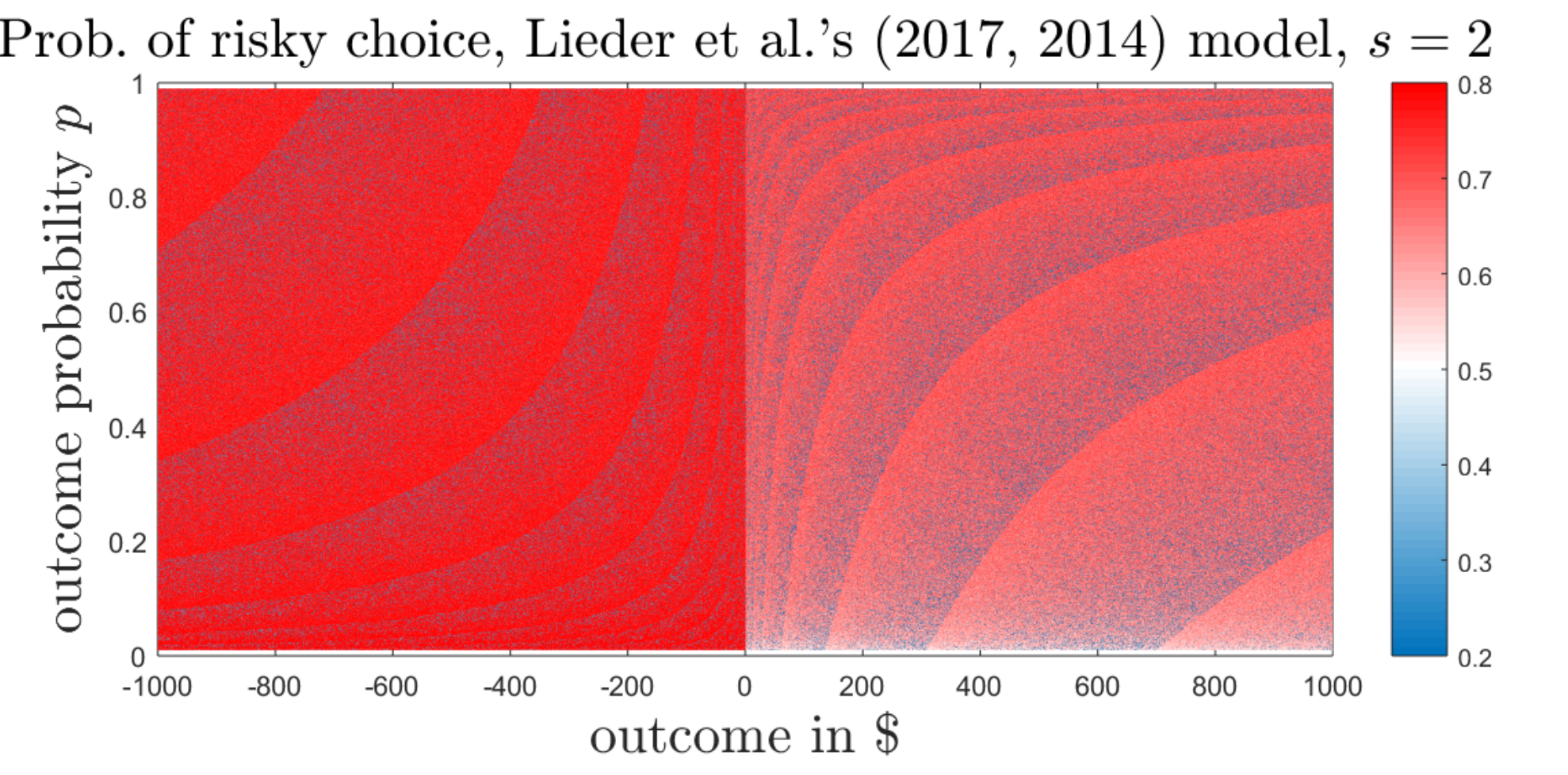} }}
    \hspace*{1pt}
    \linebreak
    \caption{\small{Accounting for the fourfold pattern of risk preferences in outcome probability (Tversky \& Kahneman, 1992), with few samples ($s=2$) and the utility function given in (\ref{util_KT}) based on the prospect theory (Tversky \& Kahneman, 1992). \textbf{(a)} Our metacognitively-rational model can account for the fourfold pattern of risk preferences in outcome probability, with $s=2$ and the utility function given in (\ref{util_KT}). \textbf{(b)} Lieder et al.'s (2014, 2017) cognitively-rational model prediction for the probability of choosing the risky choice, with $s=2$ and the utility function given in (\ref{util_KT}).}}
    \label{fig_4fold_KT_combined}
\end{figure*}

\setcounter{figure}{0}
\begin{figure}[h!]
\centering
\includegraphics[trim = 15mm 69mm 20mm 75mm, clip, width=0.3\textwidth]{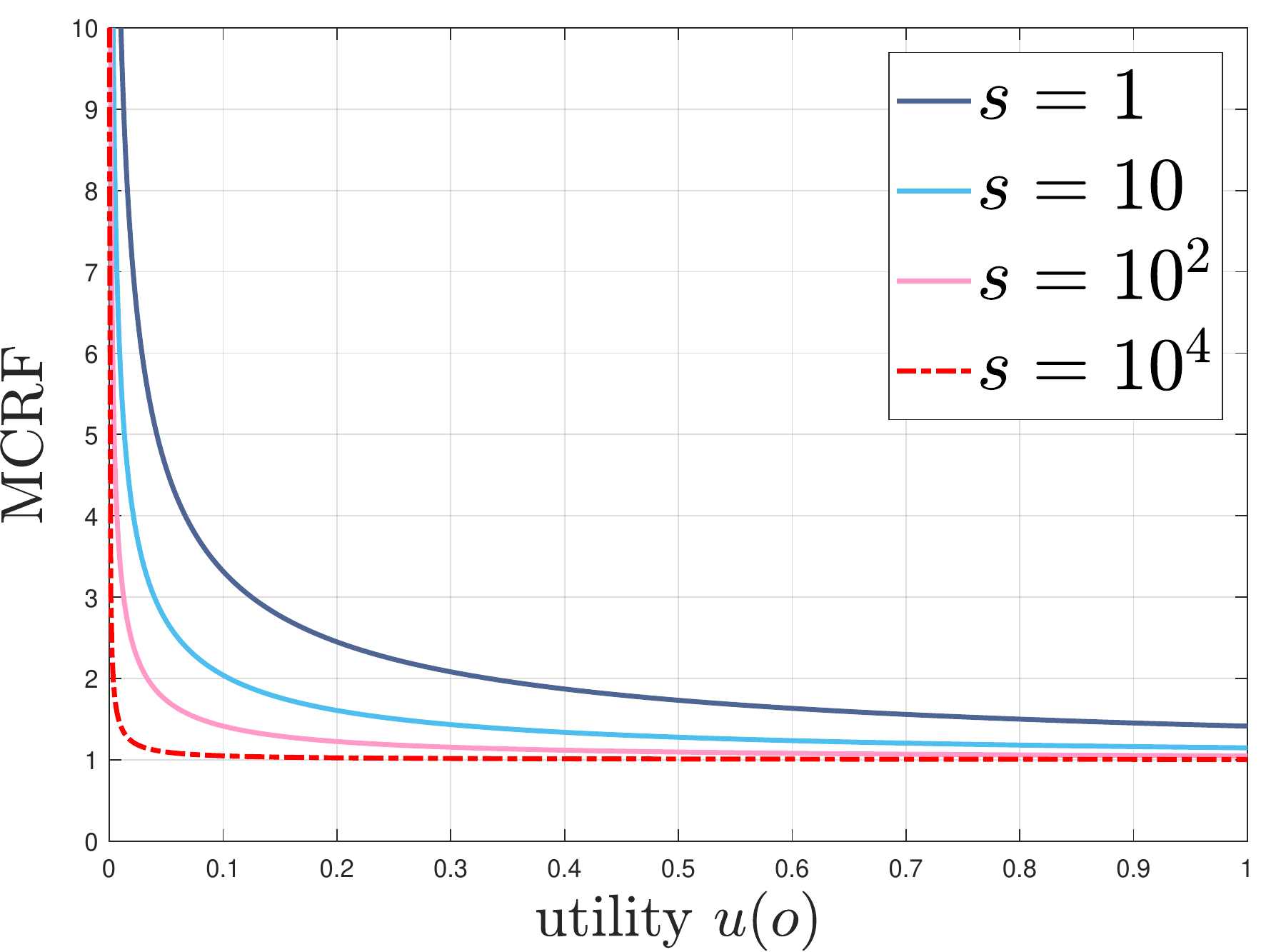}
\caption{\small{A {meta}cognitively-rational agent should over-represent extreme events precisely according to Lieder et al.'s (2014, 2017) cognitively-rational model, evidenced by the curves converging to 1 as $u(o)\rightarrow+\infty$. Importantly, however, a metacognitively-rational agent should also over-represent mundane events significantly more than what a merely cognitively-rational model prescribes, evidenced by the curves overshooting at the neighborhood of $u(o)=0$.}}
\label{fig_MCRF}
\end{figure}

\section{Extreme Eventualities are Over-represented Regardless of their Frequency}
A simple investigation of the metacognitively-rational importance distribution $q^\ast_{meta}$ given in (\ref{eq_meta_rational_q}) yields an important prediction of our model:~Extreme eventualities should be over-represented in decision-making, regardless of how rare or frequent they are. Importantly, this effects is already empirically confirmed (Lieder et al., 2017). Note that this coverage is to be expected as our proposed model subsumes the model outlined in Lieder et al. (2014, 2017).

Importantly, a detailed analysis of MCRF reveals that a \emph{meta}cognitively-rational agent should over-represent extreme events precisely according to Lieder et al.'s (2014, 2017) cognitively-rational model, however, it should also over-represent mundane events significantly more than what the cognitively-rational model by Lieder et al.~prescribes. These findings are depicted in Fig.~\ref{fig_MCRF}.

Next, we formally show that when the number of samples $s$ is sufficiently large (i.e., for large sample size regime), our  proposed metacognitively-rational importance distribution $q^\ast_{meta}$ converges to the cognitively-rational importance distribution of Lieder et al.~(2014, 2017) given in (\ref{eq_lieder_q}).\footnote{More accurately, in formal terms, $q^\ast_{meta}$ converges to (\ref{eq_lieder_q})\linebreak \emph{almost~surely}, except at $u(o)=0$. Notice that despite the unboundedness of MCRF at $u(o)=0$ (see Fig.~\ref{fig_MCRF}), $q^\ast_{meta}$ remains bounded at $u(o)=0$.}

\textbf{Proposition~1.} \emph{When the number of mental simulations $s$ is large, $q^\ast_{meta}$ converges to the importance distribution given in (\ref{eq_lieder_q}). Formally, assuming $u(o)-\mathbb{E}_p[u(o)]\neq 0$,
\begin{eqnarray}
\lim_{s\rightarrow+\infty} q^\ast_{meta} = \dfrac{1}{Z}\ p(o)|u(o)-\mathbb{E}_p[u(o)]|,
\end{eqnarray}
where $Z$ is a normalizing constant (aka partition function).}

For a formal proof of Proposition~1,  the reader is referred to Sec.~A-II of the Appendix.

Proposition~1 formally establishes that our metacognitively rational model of Availability bias serves as a rational, meta-level model for the work by Lieder et al.~(2017, 2014), with our model converging to Lieder et al.'s when the number of samples $s$ is large. Note that, since Lieder et al.'s importance distribution was specifically derived under the assumption that $s$ is large, the result presented in Proposition~1 intuitively makes sense, and, importantly, attests to the claim that our metacognitively-rational model subsumes Lieder et al.'s cognitively-rational model, with the rationality of our model holding for \emph{both} small and large $s$'s while that of Lieder et al.'s only for large $s$'s. 

\section{Framing Effect in Decision-Making}
Past work has documented that people's risk preferences are inconsistent and context-dependent (see e.g., Tversky \& Kahneman, 1992; Markovitz, 1952). For example, in choosing between a safe gamble (low payoff with high probability) and a risky gamble (high payoff with low probability), risk preferences change depending on the probabilities of the gambles (Tversky \& Kahneman, 1992), the amount offered (Markovitz, 1952), and whether those gambles are framed as a gain or loss (Tversky \& Kahneman, 1992).

In what follows, we show that our metacognitively-rational model can account for two well-known framing effects in human decision-making under risk: the fourfold pattern of risk preferences in outcome probability (Tversky \& Kahneman, 1992) and in outcome magnitude (Markovitz, 1952). Thus, our model establishes the first metacognitively-rational basis for those effects.

\subsection{Fourfold Pattern of Risk Preferences in Outcome Probability}
Framing outcomes as losses rather than gains can reverse people's risk preferences (Tversky \& Kahneman, 1992): In the domain of gains people prefer a lottery ($o$ dollars with probability $p$) to its expected value (i.e., risk seeking) when $p<0.5$, but when $p>0.5$ they prefer the expected value (i.e., risk-aversion). Conversely, in the domain of losses people are risk seeking when $p<0.5$, and risk averse when $p>0.5$. This phenomenon is known as the fourfold pattern of risk preferences in probability outcome. Next we show that our metacognitively-rational model can simulate this effect. Following the prescriptions of the \emph{prospect theory} (Tversky \& Kahneman, 1992), as did Lieder et al.~(2014) postulate, we assume that the agent's utility function can be modeled by:
\begin{eqnarray}
u(o)=\left\{
\begin{array}{ll}
o^{0.85}&\quad \text{if $o$}\geq 0,\\
{-}|o|^{0.95}&\quad \text{if $o<0$}.
\end{array}
\right.
\label{util_KT}
\end{eqnarray}  

Normatively, people should make their choice depending on whether the expected value of the utility difference $\Delta u(o)$ is negative or positive:
\begin{eqnarray}
\Delta u(o)=\left\{
\begin{array}{ll}
u(o)-u(p\times o)&\quad \text{with probability $p$},\\
-u(p\times o)&\quad \text{with probability $1-p$}.
\end{array}
\right.
\end{eqnarray}

Fig.~\ref{fig_4fold_KT_combined}(a) shows that our metacognitively-rational model can account for the fourfold pattern of risk preferences in outcome probability (Tversky \& Kahneman, 1992), with the utility function given in (\ref{util_KT}) based on the prospect theory (Tversky \& Kahneman, 1992) and very few samples ($s=2$). This result is fully consistent with past work suggesting that people often use very few samples in probabilistic inference and reasoning under uncertainty (e.g., Vul et al., 2014; Battaglia et al. 2013; Lake et al., 2017; Gershman, Horvitz, \& Tenenbaum, 2015; Hertwig \& Pleskac, 2010; Griffiths et al., 2012; Gershman, Vul, \& Tenenbaum, 2012; Bonawitz et al., 2014).

Fig.~\ref{fig_4fold_KT_combined}(b) shows Lieder et al.'s (2014, 2017) cognitively-rational model prediction for the probability of choosing the risky choice, with $s=2$ and the utility function given in (\ref{util_KT}) based on the prospect theory (Tversky \& Kahneman, 1992). Lieder et al.'s cognitively-rational model seems unable to account for the \emph{probability of risky choice} suggested by Tversky and Kahneman (1992) using a suggested utility function by the prospect theory given in (\ref{util_KT}); our simulations suggest that this apparent failure also holds for other values of $s$. However, Lieder et al.'s (2014, 2017) cognitively-rational model can partially account for this effect (see Fig.~\ref{fig_4fold_expect_imp_sampling_estimator}) based on the expected value of the importance sampling estimator given in (\ref{eq_norm_is_estimator}), $\mathbb{E}[\hat{E}]$, replicating the finding reported in Lieder et al.'s (2014) Fig~3.

\setcounter{figure}{2}
\begin{figure}[h!]
\centering
\includegraphics[trim = 5mm 90mm 10mm 91mm,clip,width=0.45\textwidth]{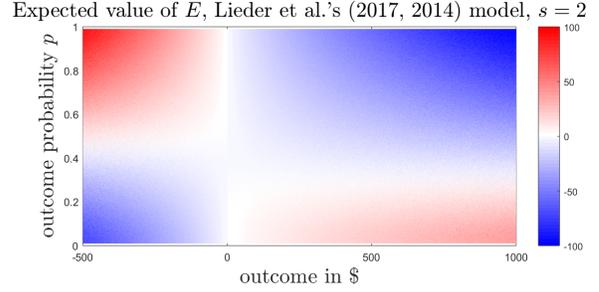}
\caption{\small{Expected value of the importance sampling estimator given in (\ref{eq_norm_is_estimator})}, $\mathbb{E}[\hat{E}]$, with $s=2$ and and the utility function given in (\ref{util_KT}), showing that Lieder et al.'s (2014, 2017) cognitively-rational model can partially account for the fourfold pattern of risk preferences in outcome probability. This replicates the finding reported in Lieder et al.'s (2014) Fig~3. However, Lieder et al.'s (2017, 2014) model appears to be unable to account for the \emph{probability of risky choice} suggested by Tversky and Kahneman (1992), using the utility function given in (\ref{util_KT}) based on the prospect theory; cf.~Fig.~\ref{fig_4fold_KT_combined}(b).}
\label{fig_4fold_expect_imp_sampling_estimator}
\end{figure}

In their recent work, Lieder et al.~(2017) showed that their cognitively-rational model can better account for the for the fourfold pattern of risk preferences in outcome probability, provided that the utility function is noisy (\emph{efficient neural coding}, Summerfield and Tsetsos, 2015); see Fig.~4 in Lieder et al.~(2017).\footnote{Specifically, Lieder et al. (2017) adopt the noisy utility function $u(o)=\frac{o}{o_{max}-o_{min}}+\epsilon,$ where $\epsilon$ is an additive Gaussian noise, i.e., $\epsilon\sim N(0,\sigma^2)$.} The result reported in Fig.~\ref{fig_4fold_KT_combined}(a) strongly suggests that this effect can be accounted for by a purely metacognitively-rational model together with a utility function fully consistent with the prospect theory (Tversky \& Kahneman, 1992), without necessarily having to invoke a noisy utility function (see Lieder et al., 2017, Appendix C).

\subsection{Fourfold Pattern of Risk Preferences in Outcome Magnitude}
\begin{figure*}[h!]
    \centering
    \subfloat[]{{\includegraphics[trim = 15mm 40mm 30mm 40mm,clip,width=0.57\textwidth]{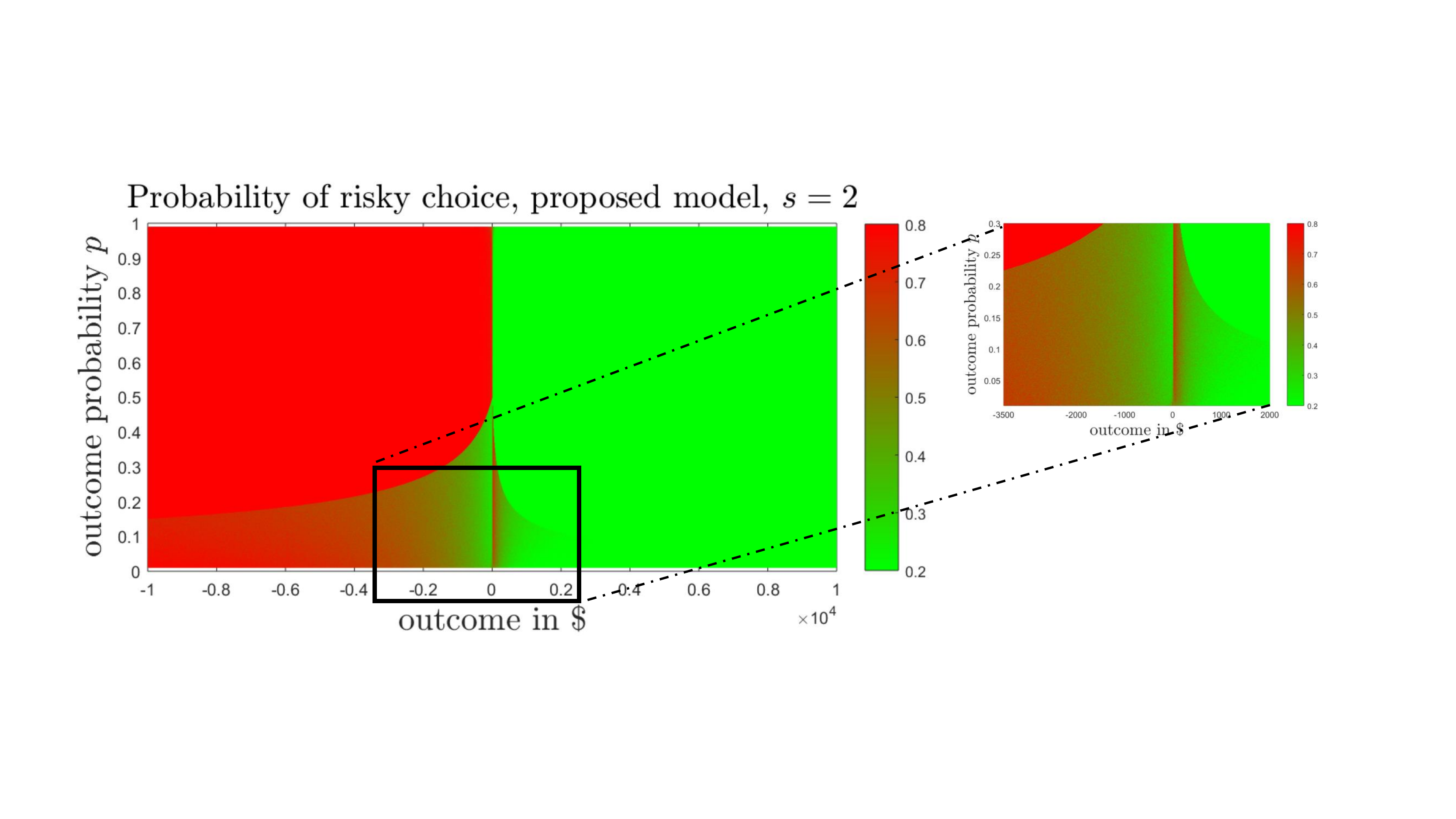} }}
    \hspace*{1pt}
    \subfloat[]{{\includegraphics[trim = 5mm 90mm 10mm 90mm,clip,width=0.43\textwidth]{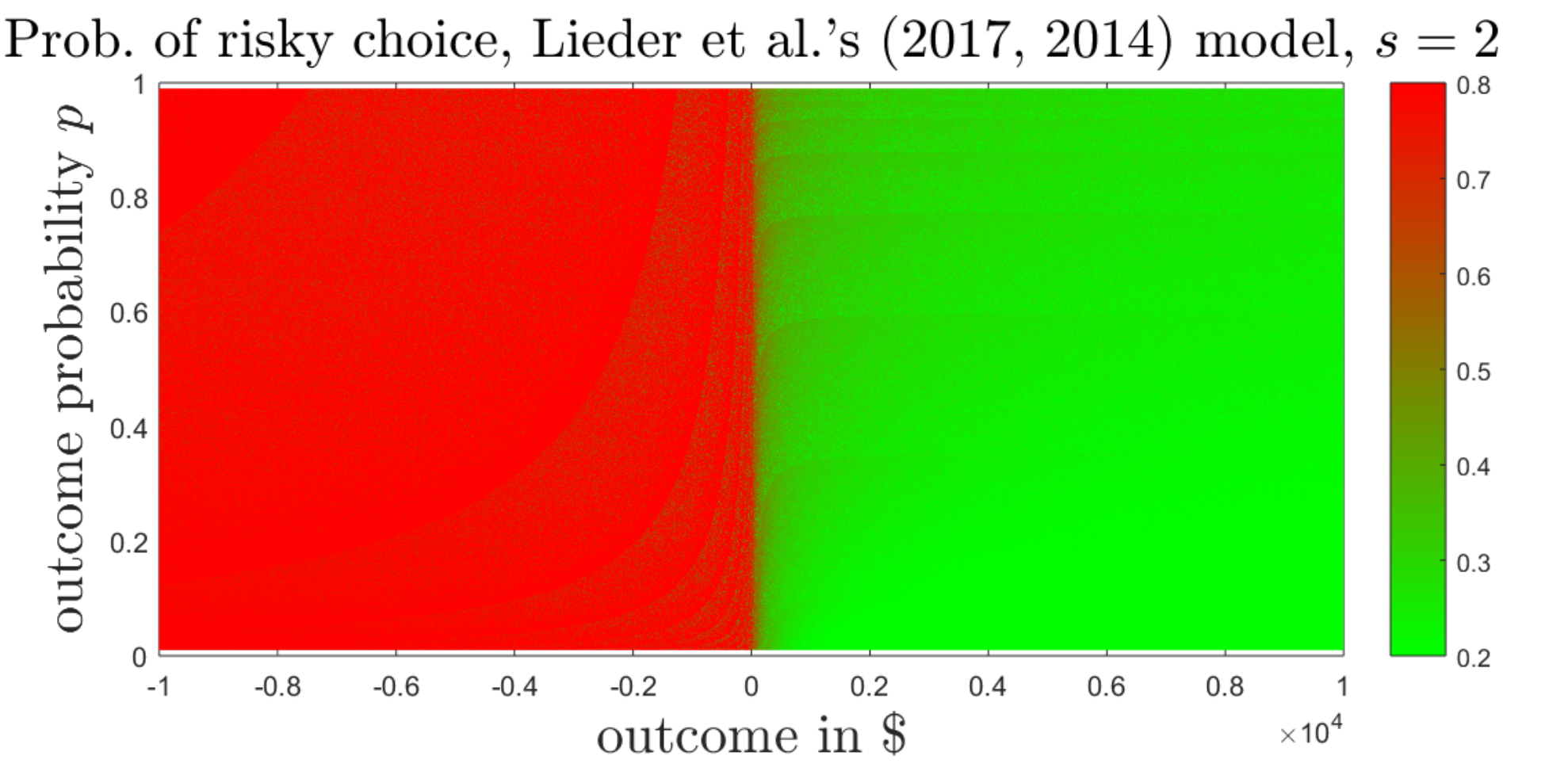} }}
    \hspace*{1pt}
    \linebreak
    \caption{\small{Simulating the fourfold pattern of risk preferences in outcome magnitude, with few samples $(s=2)$ and the normalized logarithmic utility function in (\ref{eq_norm_log_util}) with $\alpha=0.032$ and $\beta=0.0031$. \textbf{(a)} Our metacognitively-rational model can account for this effect:~Moving from left to right along the $x$-axis within the boxed region clearly shows the risk preference reversal from risk-seeking to risk-aversion (in losses), back to risk-seeking and finally to risk-aversion (in gains). For ease of visualization, a magnified version of the part lying within the black square is shown on the top-right. \textbf{(b)} Lieder et al.'s (2014, 2017) cognitively-rational model prediction under the same setting as \textbf{(a)}.}}
    \label{fig_4fold_Markowitz_combined}
\end{figure*}

Past work in behavioral economics has documented another curious inconsistency in human decision making under risk: the fourfold of risk preferences in outcome magnitude (Markovitz, 1952; Hershey \& Schoemaker, 1980; Scholten \& Read, 2014). Concretely, in choosing between a sure thing and a low-probability risky gamble people demonstrate the following behavioral pattern: In moderate-to-large outcomes, people are risk-averse for gains and risk-seeking for losses. This pattern reverses when outcomes are small, with people being risk-seeking for gains and risk-averse for losses. For example, people would rather choose a sure $1$ million dollar option rather than a (low-probability) risky gamble yielding $\$ 10$ million dollars with probability $0.1$ and nothing otherwise (Hershey \& Schoemaker, 1980). When framed in the context of losses, people prefer a risky gamble yielding $\$ 10$ million dollar loss with probability $0.1$ and nothing otherwise, over a sure loss of $\$ 1$ million dollars (Markowitz, 1952).

The prospect theory (Kahneman \& Tversky, 1979; Tversky \& Kahneman, 1992), in its original form, cannot account for the fourfold of risk preferences in outcome magnitude (Scholten \& Read, 2014). However, Scholten and Read (2014) show that, armed with a particular choice of utility function, the prospect theory can accommodate this effect. Concretely, they show that the prospect theory can best account for this effect by adopting the \emph{normalized logarithmic} utility function (Rachlin 1992; Scholten \& Read, 2010; Kirby, 2011; Kontek, 2011):
\begin{eqnarray}
u_{nlog}(o)=\left\{
\begin{array}{ll}
\dfrac{1}{\alpha}\log(1+\alpha .o)\quad &\text{if $o\geq 0$},\\
-\dfrac{\lambda}{\beta}\log(1-\beta .o)\quad &\text{if $o<0$},
\end{array}
\right.
\label{eq_norm_log_util}
\end{eqnarray}
where $\alpha,\beta\in\mathbb{R}^{> 0},\ \lambda\geq 1$ are free parameters.

Using empirical data, Scholten and Read (2014) found the maximum-likelihood estimates of $\alpha$ and $\beta$ to be $0.032$ and $0.0031$, respectively (see Scholten and Read, 2014, Table 4). Adopting the normalized logarithmic utility function in (\ref{eq_norm_log_util}) with $\lambda=1, \alpha=0.032,\beta=0.0031$, we show that our model can account for the fourfold pattern of risk preferences in outcome magnitude (see Fig.~\ref{fig_4fold_Markowitz_combined}(a)). However, Lieder et al.'s (2014, 2017) cognitively-rational model appears to be unable to account for this effect under the same setting. Again, our simulations suggest this apparent failure holds for other values of $s$. These findings suggest that the fourfold pattern of risk preferences in outcome magnitude could stem from the optimization of a boundedly-rational agent's  decision strategy at the metacognitive level, as suggested by (\ref{eq_meta_rational_q}).

\section{Sensitivity Analysis}
As discussed earlier, a metacognitively-rational agent optimizes their decision strategy (in our case, their importance distribution for mental simulations) according to time availability. This requires the agent to have a good estimate of the number of samples $s$ they will likely draw within the available time frame, using which they can appropriately select their importance distribution $q_{meta}^\ast$. However, a crucial question immediately presents itself: What if the agent is inaccurate at approximating the number of samples they get to draw before making their decision? After all, it seems plausible to assume that the agent would only have a rough estimate of the parameter $s$. Thus, it would be very likely that there would be a mismatch between the number of samples the agent thinks they can draw, and the actual number of samples they finally draw. Our model nicely allows for a quantitative investigation of the effects of such a mismatch. The parameter $s$ in (\ref{eq_meta_rational_q}) indicates the the number of samples the agent \emph{thinks} they can draw, whereas the parameter $s$ in (\ref{eq_norm_is_estimator}) reflects the the number of samples the agent actually \emph{draws} before making a decision. It is worth noting that the cognitively-rational model by Lieder et al. (2014, 2017) does not permit the investigation of the possible mismatch alluded to above, as the parameter $s$ does not feature in Lieder et al.'s importance distribution (Eq.~(\ref{eq_lieder_q})).

Intriguingly, our model demonstrates a striking insensitivity to such mismatches: Even if the the number of sample the agent {thinks} they can draw is unimaginably greater (to be precise, $10^8$ times greater) than the the number of samples they actually draw before making their decision, the agent should still show the fourfold patterns. Figures are omitted due to lack of space.

\section{General Discussion}
People overestimate the probability of extreme events, and show the fourfold pattern of risk preferences in outcome probability (Tversky \& Kahneman, 1992) and in outcome magnitude (Markovitz, 1952) in decision-making under risk; these effects are generally taken as evidence against human rationality. In this work, we presented the first metacognitively-rational process model which can account for those effects, appearing to suggest that they might not be signs of human irrationality after all, but the result of a boundedly-rational decision-maker \emph{optimizing} their decision strategy (in our case, their importance sampling distribution for performing mental simulations) in accord with time availability. In fact, it can be shown that the metacognitively-rational importance distribution $q_{meta}^\ast$ in (\ref{eq_meta_rational_q}) allows the decision-maker to ensure an upper-bound on the MSE of their estimator for the expected value in (\ref{eq_exp_utility}) using \emph{minimal} number of samples, thereby demonstrating signs of {economy} (\emph{rational minimalist program,} Nobandegani, 2017). Furthermore, our model is unimaginably robust to inaccurate estimations its focal parameter $s$, positioning it as the first rational process model we know of which scores near-perfectly in optimality, economical use of limited cognitive resources, and robustness, all at the same time.

The metacognitively-rational process model presented in this work and Lieder et al.'s (2014, 2017) cognitively-rational process model seem to suggest that a (boundedly) rationalist approach to understanding human decision-making at the algorithmic level might be a fruitful endeavor. In fact, the influential Rescorla-Wagner model (1972) and its extension temporal-difference learning model (Sutton \& Barto, 1987; Sutton \& Barto, 1998) can be given solid rational grounds based on linear-Gaussian generative models and the Kalman filtering paradigm, a rational scheme in signal detection theory (Kalman, 1960).

Our model also makes a counterintuitive (normative) prediction, which, if confirmed, would have surprising implications for human decision-making under risk: In choosing between a lottery ($o$ dollars with probability $p$) and its expected value ($p\times o$), people should qualitatively behave the same under the following two conditions: (i)~making a decision based on a mere single sample (i.e.,~under extremely high time pressure) and (ii)~making a decision based on a great many samples (i.e.,~after a along deliberation time).  Note that, given the normative status of our model, this is exactly the behavior that a boundedly-rational agent should manifest, a finding which would be of great interest for the artificial intelligence community. If confirmed, this prediction seems to suggest an intriguing possibility for human decision-making under risk: people's performance after long deliberation times is qualitatively similar to their performance under extremely high time pressure (i.e., $s=1$). This clearly serves as a motivation for avoiding over-thinking.

For their cognitively-rational process model, Lieder et al. (2017) proposed a neurally-plausible learning mechanism , a simple modifications of which permits our metacognitively-rational to be learned in a neurally-plausible manner as well. Lieder et al.~(2017) showed that their model can account for impressively wide range of cognitive biases in decisions from experience, decisions from description, and memory recall. Future work should investigate how well our metacognitively-rational model can account for those biases. The fact that our model subsumes Lieder et al.'s (2014, 2017) model (see Proposition~1), greatly elevates the possibility of our model capturing those effects as well.

To our knowledge, our model is the first \emph{metacognitive}, resource-rational process model of cognitive biases, and generally sheds light on possible rational grounds of human decision-making. We hope to have made some progress in this exciting direction.

\vspace*{5pt}
\hspace*{-15pt} \textbf{Acknowledgments:}~\small{We would like to thank Falk Lieder for fruitful discussions. This works is supported by an operating grant to TRS from Natural Sciences and Engineering Research Council of Canada.}\\

\nocite{markowitz1952utility}
\nocite{tversky1992advances}
\nocite{kahneman1972subjective}
\nocite{maule1993theoretical}
\nocite{svenson1993time}
\nocite{lichtenstein1978judged}
\nocite{cruciani2011positive}
\nocite{rothman1996absolute}
\nocite{sunstein2011overreaction}
\nocite{lieder2014high}
\nocite{lieder2017overrepresentation}
\nocite{von1955theory}
\nocite{hammersley1968monte}
\nocite{fiser2010statistically}
\nocite{vul2014one}
\nocite{denison2013rational}
\nocite{griffiths2006optimal}
\nocite{geweke1989bayesian}
\nocite{shi2009neural}
\nocite{shi2010exemplar}
\nocite{hershey1980prospect}
\nocite{scholten2014prospect}
\nocite{rachlin1992diminishing}
\nocite{scholten2010psychology}
\nocite{kirby2011empirical}
\nocite{kontek2011mental}
\nocite{Nobandegani2017phdthesis}
\nocite{rescorla1972theory}
\nocite{sutton1987temporal}
\nocite{sutton1998reinforcement}
\nocite{kalman1960new}
\nocite{poor2013introduction}
\nocite{battaglia2013simulation}
\nocite{lake2017building}
\nocite{gershman2015computational}
\nocite{hertwig2010decisions}
\nocite{griffiths2012bridging}
\nocite{gershman2012multistability}
\nocite{bonawitz2014probabilistic}

\section*{Appendix}

\subsection*{A-I:\hspace*{5pt} Proof of $q_{meta}^\ast$ Given in (\ref{eq_meta_rational_q})}
Using (4)-(5), the mean-squared error (MSE) of the (normalized) importance sampling estimator $\hat{E}$ (Eq.~3) can be written as:
\begin{eqnarray*}
\mathbb{E}[(\hat{E}-\mathbb{E}[u(o)])^2] &\approx & \dfrac{1}{s} \int \dfrac{p(o)^2}{q(o)}(\mathbb{E}_p[u(o)]-u(o))^2 do + \\ 
&&\dfrac{1}{s^2} \big[\int \dfrac{p(o)^2}{q(o)}(\mathbb{E}_p[u(o)]-u(o)) do\big]^2
\end{eqnarray*}

Under the mild technical condition
\begin{eqnarray*}
\dfrac{1}{s} \int \dfrac{p(o)^2}{q(o)}(\mathbb{E}_p[u(o)]-u(o))^2 do \leq \big[\dfrac{1}{\sqrt{s}} \int \dfrac{p(o)^2}{q(o)}(\mathbb{E}_p[u(o)]-u(o))^2 do \big]^2,
\end{eqnarray*}
the following holds

\begin{align*}
\mathbb{E}[(\hat{E}-\mathbb{E}[u(o)])^2] &\leq& &\bigg[\dfrac{1}{\sqrt{s}} \int \dfrac{p(o)^2}{q(o)}(\mathbb{E}_p[u(o)]-u(o))^2 do + \\
&& &\dfrac{1}{s} \bigg|\int \dfrac{p(o)^2}{q(o)}(\mathbb{E}_p[u(o)]-u(o)) do\bigg|  \bigg]^2\\
&\leq& &\bigg[\dfrac{1}{\sqrt{s}} \int \dfrac{p(o)^2}{q(o)}(\mathbb{E}_p[u(o)]-u(o))^2 do + \\
&& &\dfrac{1}{s} \int \dfrac{p(o)^2}{q(o)}\bigg|\mathbb{E}_p[u(o)]-u(o)\bigg| do  \bigg]^2
\end{align*}

Next, we show that
\begin{align*}
q_{meta}^\ast=\arg\min_{q} \bigg[\dfrac{1}{\sqrt{s}} \int \dfrac{p(o)^2}{q(o)}(\mathbb{E}_p[u(o)]-u(o))^2 do + \\
\dfrac{1}{s} \int \dfrac{p(o)^2}{q(o)}\bigg|\mathbb{E}_p[u(o)]-u(o)\bigg| do  \bigg]. \tag{A.1}
\end{align*}

Forming the Lagrangian,
\begin{eqnarray*}
L(q)&=&\dfrac{1}{\sqrt{s}} \int \dfrac{p(o)^2}{q(o)}(\mathbb{E}_p[u(o)]-u(o))^2 do + \\
&&\dfrac{1}{s} \int \dfrac{p(o)^2}{q(o)}\bigg|\mathbb{E}_p[u(o)]-u(o)\bigg| do +\\
&&\lambda(\int q(o)do -1).
\end{eqnarray*}
Equating the first variation to zero straightforwardly implies
\begin{align*}
\dfrac{\delta}{\delta q}  L(q)=0\ &\Longrightarrow  \\
&q^\ast_{meta}\propto p(o)|u(o)-\mathbb{E}_p[u(o)]| \sqrt{\dfrac{1+|u(o)-\mathbb{E}_p[u(o)]|\sqrt{s}}{|u(o)-\mathbb{E}_p[u(o)]|s}}.
\end{align*}
Note that, as a distribution (which integrates to one over all events $o$), $q^\ast_{meta}$ is invariant under any multiplicative re-scaling by a purely function of $s$, $f(s)$, which does not involve $o$. Hence, using $f(s)=\sqrt[4]{s}$, we have:
\begin{eqnarray*}
q^\ast_{meta}\propto p(o)|u(o)-\mathbb{E}_p[u(o)]| \sqrt{\dfrac{1+|u(o)-\mathbb{E}_p[u(o)]|\sqrt{s}}{|u(o)-\mathbb{E}_p[u(o)]|\sqrt{s}}},
\end{eqnarray*}
thereby granting the validity of the expression (\ref{eq_meta_rational_q}) in the main text. Finally, using Jensen's inequality, it is straightforward to show that $q^\ast_{meta}$ indeed satisfies (A.1). This completes the proof. \hfill $\blacksquare$

\subsection*{A-II:\hspace*{5pt} Proposition~1}
\emph{\textbf{Proof.}} The distribution $q^\ast_{meta}$ given in (\ref{eq_meta_rational_q}) can be re-written as:
\begin{eqnarray*}
q^\ast_{meta} \propto p(o)|u(o)-\mathbb{E}_p[u(o)]| \sqrt{1+\textcolor{blue}{\dfrac{1}{|u(o)-\mathbb{E}_p[u(o)]|\sqrt{s}}}}.
\end{eqnarray*}
Assuming $u(o)-\mathbb{E}_p[u(o)]\neq 0$, as $n\rightarrow+\infty$, the term shown in blue approaches zero. This completes the proof. \hfill $\blacksquare$

\renewcommand\bibliographytypesize{\normalsize}
\bibliographystyle{apacite}
\bibliography{ref_meta_availability}
\end{document}